\documentclass[prb,aps,twocolumn,draft,tightenlines,showpacs,floatfix,superscriptaddress]{revtex4}
\usepackage{psfig}
\usepackage{amssymb}
\begin{document}
\draft
\tighten
\title{Spin dephasing in $n$-typed GaAs quantum wells in the presence of high magnetic fields in Voigt configuration}
\author{M. Q. Weng}%
\affiliation{Structure Research Laboratory, University of Science \& %
Technology of China, Academia Sinica,  Hefei, Anhui, 230026, China}
\affiliation{Department of Physics, University of Science \& %
Technology of China, Hefei, Anhui, 230026, China}%
\altaffiliation{Mailing Address.}
\author{M. W. Wu}%
\thanks{Author to whom correspondence should be addressed}%
\email{mwwu@ustc.edu.cn}%
\affiliation{Structure Research Laboratory, University of Science \& %
Technology of China, Academia Sinica,  Hefei, Anhui, 230026, China}
\affiliation{Department of Physics, University of Science \& %
Technology of China, Hefei, Anhui, 230026, China}%
\altaffiliation{Mailing Address.}
\date{\today}
\begin{abstract}
We perform a many-body study of the spin dephasing 
due to the D'yakonov-Perel' effect 
in $n$-typed GaAs (100) quantum wells 
under high magnetic fields in the Voigt configuration 
by constructing and numerically solving the
kinetic Bloch equations. We include all the spin conserving scattering 
such as electron-phonon, the electron-nonmagnetic impurity as
well as  the electron-electron Coulomb scattering in our
theory and investigate how the spin dephasing time (SDT) is  
affected by the initial spin polarization,
 impurity, and magnetic field.
 The dephasing obtained from our theory
contains not only that due to the  effective spin-flipping scattering
first proposed by  D'yakonov and  Perel' [Zh. Eksp. Teor. Fiz. {\bf
  60}, 1954 (1971)[Sov. Phys.-JETP {\bf 38}, 1053 (1971)]], 
but also the recently proposed many-body dephasing 
due to the inhomogeneous broadening provided by the DP term
[Wu, J. Supercond.:Incorp. Novel Mechanism {\bf 14}, 245 (2001); Wu and Ning, 
Eur. Phys. J. B {\bf 18}, 373 (2000)]. We are able to investigate the spin dephasing with extra large
spin polarization (up to 100~\%) which has not been discussed both
theoretically and experimentally. A huge anomalous resonance 
of the SDT for large spin polarizations is predicted under the high magnetic field we used. 
\end{abstract}
\maketitle

\section {Introduction}

The resent development of ultrafast nonlinear optical 
experiments\cite{damen,%
wagner,baumberg_1994_prl,baumberg_1994_prb,heberle,%
buss1,crooker_1996,crooker_1997,buss2,kikkawa1,kikkawa2,%
kikkawa3,ohno1,ohno} has stimulated immense interest in spintronics 
in semiconductors as it shows great potential 
of using the spin degree of freedom of electrons in place of/in addition to
the charge degree of freedom for device application such as qubits,
quantum memory devices, and spin transistors. 
In order to make use of the spin degree of freedom in semiconductor
spintronics, it is crucial to have a thorough understanding 
of spin dephasing mechanism.

Three spin dephasing mechanisms have been proposed 
in semiconductors:\cite{meier,aronov}
the Ellit-Yafet (EY) mechanism,\cite{yafet,elliot} the D'yakonov-Perel'
(DP) mechanism,\cite{DP} and the Bir-Aronov-Pikus (BAP)
mechanism.\cite{BAP} In the EY mechanism, the spin-orbit interaction leads
to mixing of wave functions of opposite spins. This mixing results in
a nonzero electron spin flip due to impurity and phonon scattering. The DP
mechanism is due to the spin-orbit interaction in crystals without
inversion center, which results in spin state splitting of the
conduction band at $k\not=  0$. This is equivalent to an effective
magnetic field acting on the spin, with its magnitude and orientation
depending on ${\bf k}$. Finally, the BAP mechanism is originated from
the mixing of heavy hole and light hole bands induces by spin-orbit
coupling. Spin-flip (SF) scattering of electrons by holes due to the
Coulomb interaction is therefore permitted, which gives rise to spin
dephasing. The dephasing rates of these mechanisms 
for low polarized system are calculated in
the framework of single particle approximation.\cite{meier} 
For GaAs, the EY mechanism is less effective
under most conditions, due to the large band gap and low scattering
rate for high quality samples. The BAP mechanism is important
for either $p$-doped or insulating GaAs. For $n$-doped samples,
however, as holes are rapidly recombined with electrons due to
the presence of a large number of electrons, spin
dephasing due to the regular BAP mechanism is blocked. Therefore,
the DP mechanism  (or possibly the EY mechanism under certain
conditions) is the main mechanism of spin dephasing for $n$-type GaAs.

All the above mentioned spin dephasing mechanisms are either due to
the SF scattering or treated as effective SF scattering.  In
additional to these spin dephasing mechanisms, three years ago Wu
proposed a many-body spin dephasing mechanism which is due to the
inhomogeneous broadening, such as energy dependence of
$g$-factor\cite{wu_pss_2000,wu_js_2001} and/or the momentum ${\bf k}$-dependence of
the DP term,\cite{wu_js_2001,wu_jpsj_2001,wu_ssc_2002} together with the SC
scattering.\cite{wu_pss_2000,wu_ssc_2002,wu_jpsj_2001,wu_epjb_2000,wu_js_2001} Differing from the
earlier study of the spin dephasing which comes from the SF
scattering, the spin dephasing through inhomogeneous broadening is
caused by irreversibly disrupting the phases between spin dipoles and
is therefore a many-body effect.\cite{wu_jpsj_2001,wu_epjb_2000,wu_js_2001} This many-body spin
dephasing mechanism has long been overlooked in the
literature. Recently, we further showed that this inhomogeneous
broadening effect also plays an important role in the spin
transport.\cite{weng_prb_2002,weng_jap_2003} Very recently, Bronold {\em et al.} also
discussed the spin dephasing due to the $k$-dependence of the $g$
factor.\cite{bronold}

In our recent work Ref.\ \onlinecite{c0210313}, we have performed a
many-body study of the spin dephasing due to the D'yakonov-Perel'
effect in $n$-typed GaAs (100) quantum wells for high temperatures
($\geq 120$~K) under moderate magnetic fields in the Voigt
configuration by constructing and numerically solving the kinetic
Bloch equations set up by Wu {\em et
al.}.\cite{wu_prb_2000,wu_pss_2000,wu_ssc_2002,wu_jpsj_2001,wu_epjb_2000,wu_js_2001} We include all the spin
conserving scattering such as the electron-phonon, the
electron-nonmagnetic impurity as well as the electron-electron Coulomb
scattering in our theory and investigate how the spin dephasing rate
is affected by the initial spin polarization, temperature, impurity,
magnetic field as well as the electron density.  The dephasing
obtained from our theory contains not only that due to the effective
SF scattering first proposed by D'yakonov and Perel',\cite{DP} but
also the many-body dephasing due to the inhomogeneous broadening
provided by the DP term.  We show that for the electron densities we
study, the spin dephasing rate is dominated by the many-body
effect. Equally remarkable is that we are now able to investigate the
spin dephasing with extra large spin polarization (up to 100~\%) which
has not been discussed both theoretically and experimentally. We find
a dramatic decrease of the spin dephasing rate for large spin
polarizations.  The spin dephasing time (SDT), which is defined as the
inverse of the spin dephasing rate, we get at low initial spin
polarization is in agreement with the experiment both qualitatively
and quantitatively.

In this paper, we further extend the above mentioned work to the case
of the high magnetic fields ($B\ge 60$\ T). These high fields can be
achieved experimentally by the pulsed magnetic field with the stable
duration of the field pulse being miliseconds, orders of magnitude
longer than the SDT.\cite{kulb,respaud} We will show a huge anomalous
resonance of the SDT for large spin polarizations under this high
magnetic field case. We organize the paper as follows: We briefly
present our model and the kinetic equations in Sec.\ II. Then in 
Sec.\ III(A) we investigate how the SDT changes with the variation of
the initial spin polarization. In Sec.\ III(B) we
show the magnetic field dependence of the SDT.  We present the conclusion and
summary in Sec.\ IV.

\section {Kinetic Equations}

We start our investigation from an $n$-doped (100) GaAs QW with 
well width $a$. The growth direction is assumed to be $z$-axis. 
A magnetic field {\bf B} is applied along the $x$ axis. 
Due to the confinement of the QW, the
momentum states along $z$ axis are quantized.  Therefore the electron
states are characterized by a subband index $n$ and a two dimensional
wave vector ${\bf k}=(k_x, k_y)$ together with a spin index $\sigma$.
In the present paper, the subband separation is assumed to be large
enough so that only the lowest subband is populated and the transition
to the upper subbands is unimportant. Therefore, one only needs to
consider the lowest subband. For $n$-doped samples, spin dephasing
mainly comes from the DP mechanism.\cite{DP} With the DP term
included, the Hamiltonian of the electrons in the QW takes the form:
\begin{equation}
  H=\sum_{{\bf k}\sigma\sigma^{\prime}}\biggl\{
\varepsilon_{\bf k}+\bigl[g\mu_B{\bf B}+{\bf h}({\bf k})\bigr]
\cdot{\vec{\bf \sigma}_{\sigma\sigma^{\prime}}\over 2}\biggr\}
c^{\dagger}_{{\bf k}\sigma}c_{{\bf k}\sigma^{\prime}}+H_I.
\label{eq:hamiltonian}
\end{equation}
Here $\varepsilon_{{\bf k}}={\bf k}^2/2m^{\ast}$ is the energy of
electron with wavevector ${\bf k}$ and effective mass $m^{\ast}$.
$\vec{\bf \sigma}$ are the Pauli matrices. 
In QW system, the DP term is composed of the Dresselhaus
term\cite{dress} and the Rashba term.\cite{ras,rashba} The Dresselhaus
term is due to the lack of inversion symmetry in the zinc-blende 
crystal Brillouin zone and is sometimes referred to as bulk inversion
asymmetry (BIA) term.  For the (100) GaAs QW system, it can be written
as\cite{eppen,ivch}
\begin{eqnarray}
  &&h^{\mbox{BIA}}_x({\bf k})=\gamma k_x(k_y^2-\langle k_z^2\rangle), \;
  \nonumber\\
  &&h^{\mbox{BIA}}_y({\bf k})=\gamma k_y(\langle k_z^2\rangle-k_x^2), \;
\nonumber\\
&&  h^{\mbox{BIA}}_z({\bf k})=0\ .
  \label{eq:dp}
\end{eqnarray}
Here $\langle k^2_z\rangle$ represents the average of the operator
$-({\partial\over\partial z})^2$ over the electronic state of the
lowest subband and is therefore  $(\pi/a)^2$.
$\gamma=(4/3)(m^{\ast}/m_{cv})(1/\sqrt{2m^{\ast
3}E_g})(\eta/\sqrt{1-\eta/3})$ and $\eta=\Delta/(E_g+\Delta)$, in
which $E_g$ denotes the band gap; $\Delta$ represents the spin-orbit
splitting of the valence band; $m^{\ast}$ standing for the electron mass
in GaAs; and $m_{cv}$ is a constant close in magnitude to free
electron mass $m_0$.\cite{aronov} 
Whereas the Rashba term appears if the
self-consistent potential within a QW is asymmetric along the growth
direction and is therefore referred to as structure inversion
asymmetry (SIA) contribution. Its scale 
is proportional to the interface electric
field along the growth direction. For narrow band-gap
semiconductors such as InAs, the Rashba term is the main
spin-dephasing mechanism; whereas in the wide band-gap semiconductors
such as GaAs, the Dresselhaus term is dominant. 
In the present paper, we will take only the Dresselhaus term into
consideration as we focus on the spin dephasing in GaAs QW. 
The interaction Hamiltonian $H_I$ is composed of Coulomb interaction
$H_{ee}$, electron-phonon interaction $H_{ph}$, as well as
electron-impurity scattering $H_i$. Their expressions can be found in
textbooks.\cite{haug,mahan} 

We construct the kinetic Bloch equations by the nonequilibrium Green
function method\cite{haug} as follows:
\begin{equation}
  \label{eq:bloch}
  \dot{\rho}_{{\bf k},\sigma\sigma^{\prime}}
  =\dot{\rho}_{{\bf k},\sigma\sigma^{\prime}}|_{\mbox{coh}}
  +\dot{\rho}_{{\bf k},\sigma\sigma^{\prime}}|_{\mbox{scatt}}
\end{equation}
Here $\rho_{{\bf k}}$ represents the single particle density
matrix. The diagonal elements describe the electron distribution
functions $\rho_{{\bf k},\sigma\sigma}=f_{{\bf k}\sigma}$. The
off-diagonal elements $\rho_{{\bf k},{1\over
    2}-{1\over2}}\equiv\rho_{{\bf k}}$ describe 
the inter-spin-band polarizations
(coherence) of the spin coherence.\cite{wu_prb_2000} Note that 
$\rho_{{\bf k},-{1\over 2}{1\over 2}}\equiv \rho^{\ast}_{{\bf k},{1\over
    2}-{1\over 2}}=\rho^{\ast}_{{\bf k}}$. Therefore, $f_{{\bf
    k}\pm{1\over 2}}$ and $\rho_{{\bf k}}$ are the quantities to be
determined from Bloch equations. 

The coherent parts of the equation of motion for the electron
distribution function and spin coherence  are given by 
\begin{widetext}
\begin{equation}
  \label{eq:f_coh}
  {\partial f_{{\bf k},\sigma}\over \partial t}|_{\mbox{coh}}=
-2\sigma\bigl\{[g\mu_BB+h_x({\bf k})]\mbox{Im}\rho_{{\bf k}}+h_y({\bf k})
\mbox{Re}\rho_{{\bf k}}\bigr\}
+4\sigma\mbox{Im}\sum_{{\bf q}}V_{{\bf q}}\rho^{\ast}_{{\bf k}+{\bf
    q}} \rho_{{\bf k}},
\end{equation}
\begin{equation}
  \label{eq:rho_coh}
  {\partial \rho_{{\bf k}}\over \partial t}\left |_{\mbox{coh}}\right. =
  {1\over 2}[ig\mu_B B + ih_x({\bf k}) + h_y({\bf k})]
  (f_{{\bf k}{1\over 2}}-f_{{\bf k}-{1\over 2}})
  +i\sum_{{\bf q}}V_{\bf q}\bigl[(f_{{\bf k}+{\bf q}{1\over 2}}
  -f_{{\bf k}+{\bf q}-{1\over 2}})\rho_{{\bf k}}
  -\rho_{{\bf k}+{\bf q}}(f_{{\bf k}{1\over 2}}
  -f_{{\bf k}-{1\over 2}})\bigr],
\end{equation}
\end{widetext}
respectively, 
where $V_{{\bf q}}=4\pi e^2/[\kappa_0(q+q_0)]$ is the 2D Coulomb
matrix element under static screening. 
$q_0=(e^2m^{\ast}/\kappa_0)\sum_{\sigma}f_{{\bf k}=0,\sigma}$  
and $\kappa_0$ is the static dielectric constant. 
The first term
on the right hand side (RHS) of Eq.~(\ref{eq:f_coh}) describes spin
precession of electrons under the magnetic field ${\bf B}$ as well as
the effective magnetic field ${\bf h}({\bf k})$ due to the DP
effect. The scattering terms 
$\dot{\rho}_{{\bf k},\sigma\sigma^{\prime}}|_{\mbox{scatt}}$ 
contain the contribution of the electron-nonmangetic impurity
interaction and the electron-phonon coupling as well as the
electron-electron Coulomb scattering. Their expressions can be found in
our previous paper Ref. \onlinecite{c0210313}. 
The initial conditions are taken at $t=0$ as: 
\begin{equation}
\rho_{\bf k}|_{\rm t=0} = 0 \ ,
\label{eq:rho_init}
\end{equation}
\begin{equation}
f_{{\bf k}\sigma}|_{\rm t=0} = 1/\bigl\{\exp[(\varepsilon_{\bf
  k}-\mu_{\sigma})/k_BT]+1\bigr\} \ ,
\label{eq:fk_init}
\end{equation}
where $\mu_\sigma$ is the chemical potential for spin $\sigma$. 
The condition
$\mu_{\frac{1}{2}}\neq\mu_{-\frac{1}{2}}$ gives rise to the imbalance
of the electron densities of the two spin bands. Eqs.~(\ref{eq:bloch})
%% through (\ref{eq:rho_scatt}) 
together with the initial conditions 
Eqs.~(\ref{eq:rho_init}) and (\ref{eq:fk_init}) comprise the complete
set of kinetic Bloch equations of our investigation.

\section {Numerical Results}

The kinetic Bloch equations form a set of nonlinear equations. All the
unknowns to be solved appear in the scattering terms. Specifically,
the electron distribution function is no longer a Fermi distribution
because of the existence of the anisotropic DP term ${\bf h}({\bf
  k})$. This term in the coherent parts drives the electron
distribution away from an isotropic Fermi distribution. The
scattering term attempts to randomize electrons in ${\bf
  k}$-space. Obviously, both the coherent parts  and the
scattering terms have to be solved self-consistently to obtain the
distribution function and the the spin coherence. 

We numerically solve the kinetic Bloch equations in such a
self-consistent fashion to study the spin precession between the
spin-up and -down bands. We include electron-phonon scattering and the
electron-electron interaction throughout our computation.  As we
concentrate on the relatively high temperature regime in the present
study, for electron-phonon scattering we only need to include
electron-LO phonon scattering. Electron-impurity scattering is
sometimes excluded.  As discussed in the previous
paper,\cite{wu_prb_2000,kuhn} irreversible spin dephasing can be well defined
by the slope of the envelope of the incoherently summed spin coherence
$\rho(t)=\sum_{{\bf k}}|\rho_{{\bf k}}|$.  The material parameters of
GaAs for our calculation are tabulated in
Table~\ref{table1}.\cite{made} The method of the numerical calculation
has been laid out in detail in our previous papers.\cite{wu_pss_2000,c0210313}
In the present calculation the total electron density $N_e$ is chose
to be $4\times 10^{11}$~cm$^{-2}$ which is a typical electron density
in the $n$-typed GaAs QW's. Differing from our previous paper where
the applied magnetic field $B$ is the moderate one, in this paper $B$
is chose to be $60$~T unless otherwise specified. Although this large
magnetic field maybe an unpractical one in the device application, it
is of theoretical interest to understand and study the spin dephasing
in this extreme condition. Moreover, this study can also throw lights
on the understanding of spin dephasing under a moderate magnetic field
of other materials with large Land\'e $g$-factors such as InAs where
$g=15$.

\begin{table}[htbp]
  \centering
  \begin{tabular}{lllllll}
    \hline
    $\kappa_\infty$ & \mbox{}\hspace{1.25cm} &
    10.8 & \mbox{}\hspace{1.25cm} &
    $\kappa_0$ & \mbox{}
    \hspace{1.25cm} & 12.9\\ 
    
    $\omega_0$ & & 35.4~meV & & $m^*$ & &0.067~$m_0$\\
    $\Delta$ & &0.341~eV & &$E_g$ & &1.55~eV\\
$g$&&0.44&&&&\\
\hline
  \end{tabular}
  
  \caption{Parameters used in the numerical calculations}
  \label{table1}
\end{table}

We have studied SDT under the high magnetic fields for all the
situations as we did in Ref.\ \onlinecite{c0210313}. Many features
such as the temperature dependence, electron density dependence as
well as the contribution of the Coulomb scattering are similar to the
results we got under moderate magnetic fields. In this paper, we just
report the results with different properties. Our main results are
plotted in Figs. 1 to 5.

\begin{figure}[htb]
  \psfig{figure=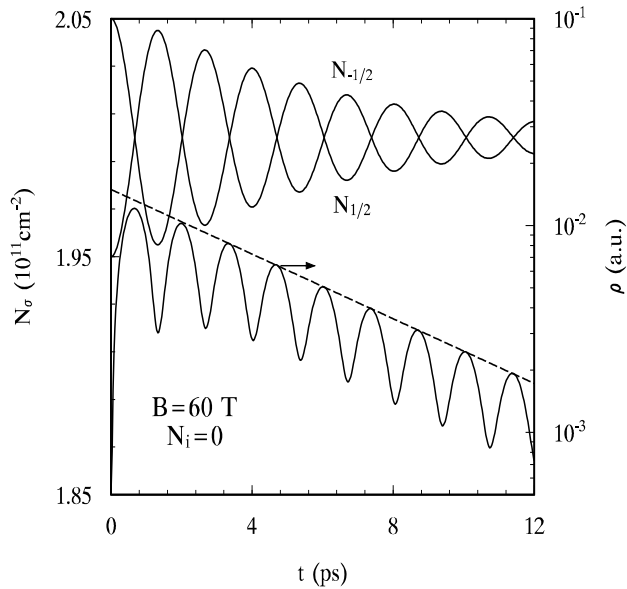,width=9.cm,height=8.5cm,angle=0}
  \caption{Electron densities of up spin and down spin
    and  the incoherently summed spin coherence $\rho$ 
    versus time $t$ for a GaAs QW with the initial spin  polarization
    $P=2.5\%$ at 
    $T=200$\ K. The dashed line gives the evelope of $\rho$.
    Note the scale of the spin coherence is on the right side of the
    figure.} 
  \label{fig1}
\end{figure}

In Fig. 1 we plot a typical temporal evolution of the spin signal in a
GaAs QW at $T=200$~K where the electron densities in the spin-up and
-down bands together with the incoherently summed spin coherence are
plotted versus time for $N_i=0$.  At $t=0$, the initial spin
polarization $P=(N_{1/2}-N_{-1/2})/(N_{1/2}+N_{-1/2})$ is $2.5\%$.  It
is seen from the figure that excess electrons in the spin-up band
start to flip to the spin-down band at $t=0$ due to the presence of
the magnetic field and the DP term ${\bf h}({\bf k})$. In the meantime
the spin coherence $\rho$ accumulates. At about $0.67$~ps, the
electron densities in the two spin bands become equal and the spin
coherence reaches its maximum. Then the spin coherence starts to feed
back and the electron density in the spin-down band exceeds that in
the spin-up band while $\rho$ deceases. At about $1.3$~ps, $\rho$
reach its minimum, while the density difference in the two spin bands
reaches its maximum again with the excess electrons now in the
spin-down band. Due to the the dephasing, the second peak is lower
than the first one. This oscillation goes on until the amplitude of
the oscillation becomes zero due to the dephasing. 
In Fig.~\ref{fig1}, we also plotted the
slope of the envelope of the incoherently summed spin coherence as
dashed line. From the slope one gets the spin dephasing time.

\subsection{Spin polarization dependence of the spin dephasing time}

We first study the spin polarization dependence of the SDT.
As our theory is a many-body one and 
we include all the scatterings, especially the
Coulomb scattering, in our calculation, we are able to calculate the 
SDT with large spin polarization. 

\begin{figure}[htb]
  \psfig{figure=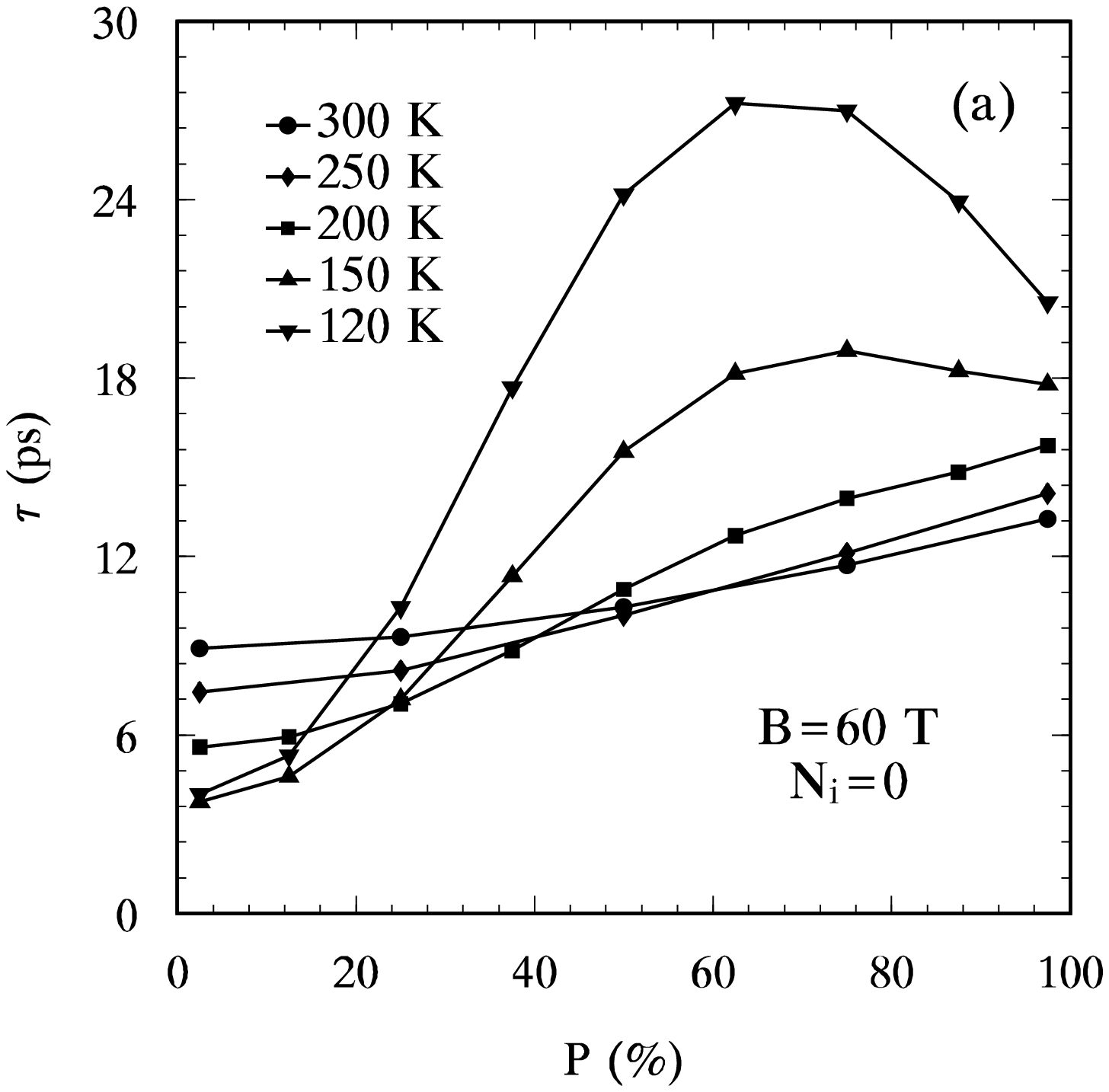,width=9.cm,height=8.5cm,angle=0}
  \psfig{figure=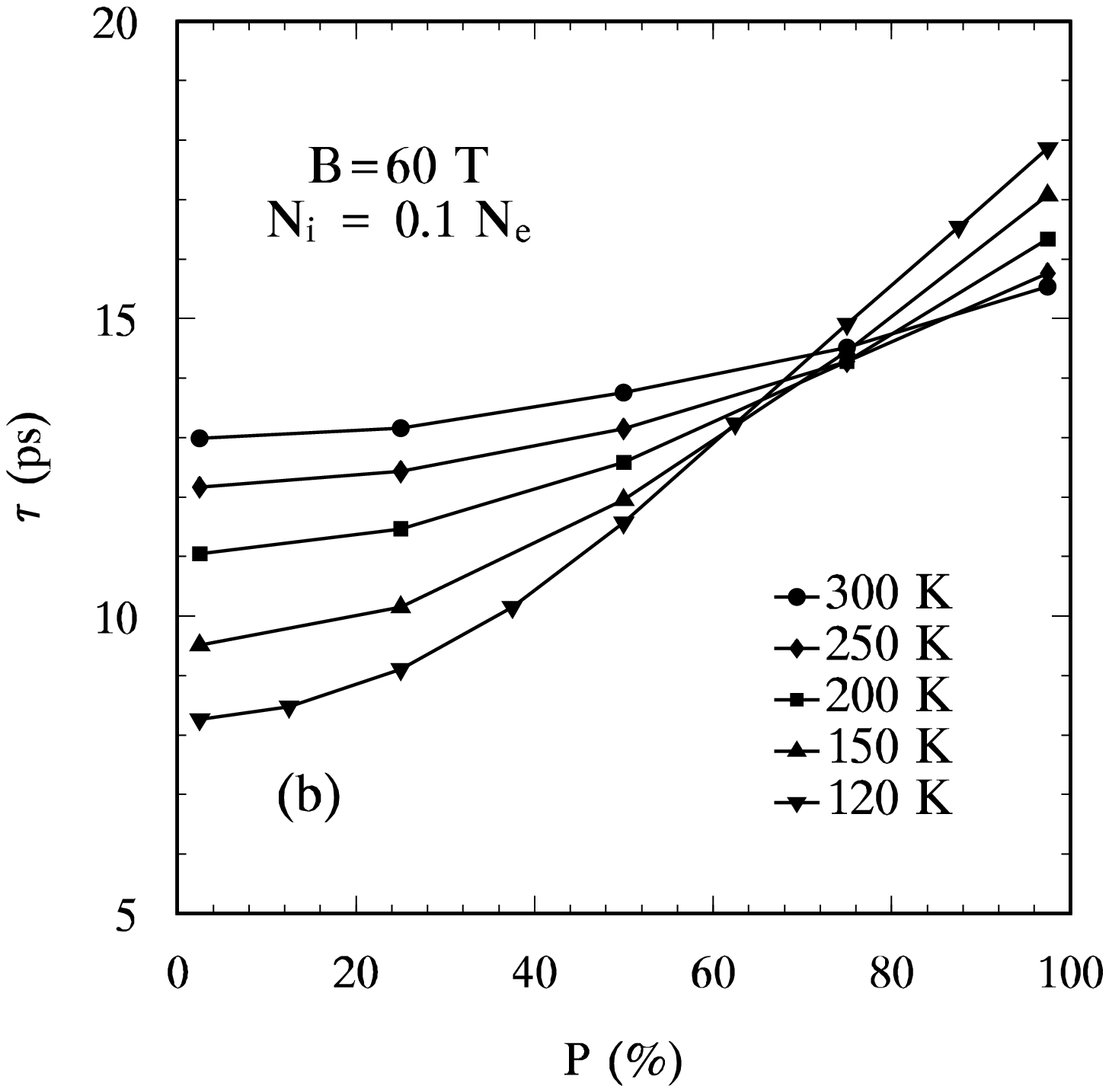,width=9.cm,height=8.5cm,angle=0}
  \caption{Spin dephasing time $\tau$ versus the initial spin
    polarization $P$ for a GaAs QW with
    different impurity concentration and different temperatures. 
    Down triangle ($\blacktriangledown$): $T=120$\ K; 
    Up triangle ($\blacktriangle$): $T=150$\ K;
    Square ($\blacksquare$): $T=200$\ K;
    Diamond ($\blacklozenge$): $T=250$\ K; 
    Circle ($\bullet$): $T=300$\ K. The impurity densities in (a) and
    (b) are 0 and $0.1 N_e$ respectively. The lines are plotted for
    the aid of eyes. 
  }
  \label{fig3}
\end{figure}

In Fig.~\ref{fig3}, SDT $\tau$ is plotted against the initial spin
polarization $P$ for GaAs QW's with $N_i=0$ [Fig.~\ref{fig3}(a)] and
$N_i=0.1 N_e$ [Fig.~\ref{fig3}(b)] at different
temperatures. Differing from the moderate magnetic field case where
the SDT increases monotonically with the initial spin
polarization,\cite{c0210313}
here the most striking feature of the impurity-free case is the huge
anomalous peaks of the SDT in low temperatures.  For $T=120$\ K, the
peak value of the SDT is about 6 times higher than that of low initial
spin polarization. It is also seen from the
figure that the anomalous 
peak is reduced with the increase of temperature and the peak shifts
to higher polarization. For $T>200$\ K there is no anomalous peak.

\begin{figure}[htb]
  \psfig{figure=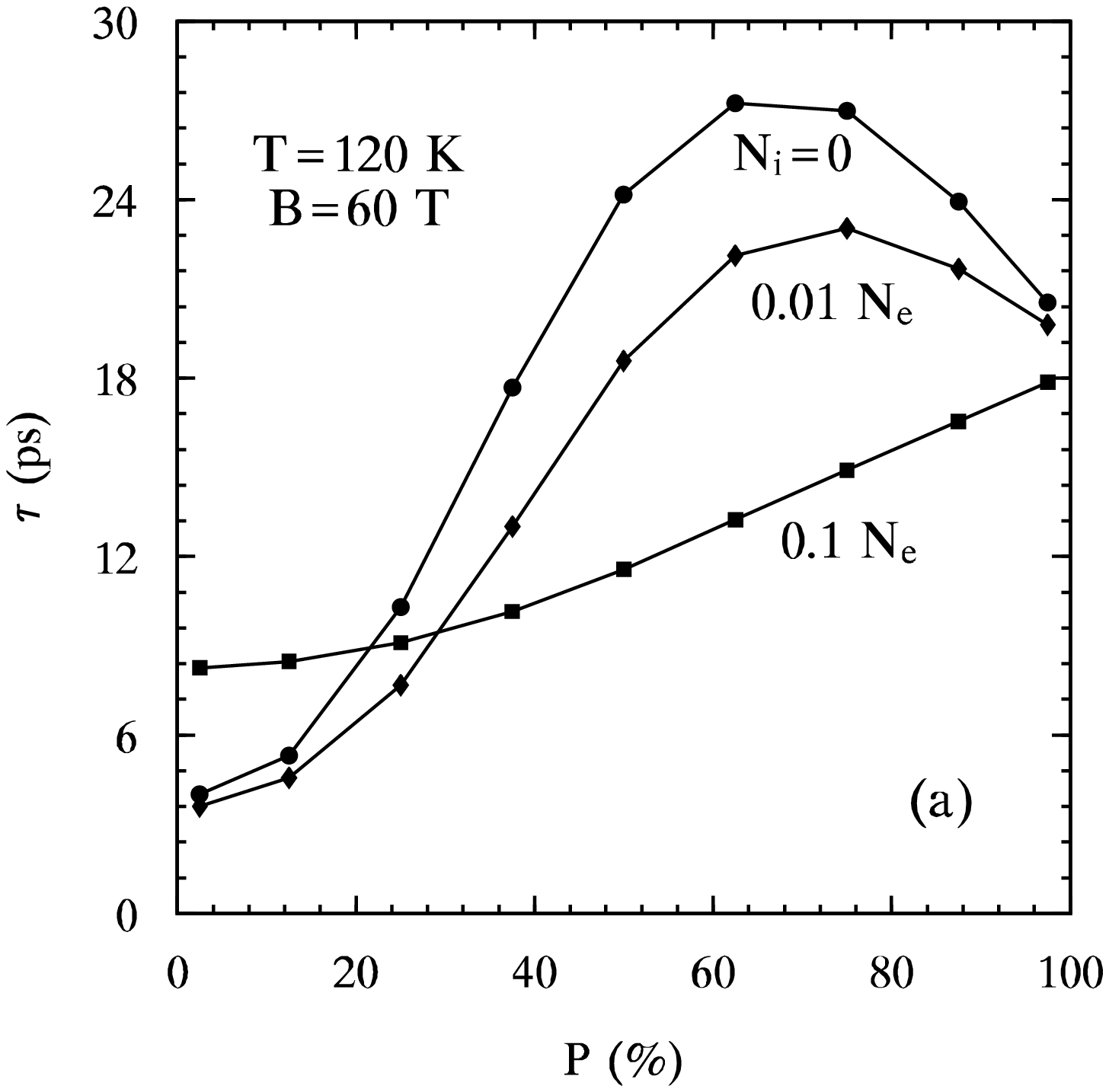,width=9.cm,height=8.5cm,angle=0}
  \psfig{figure=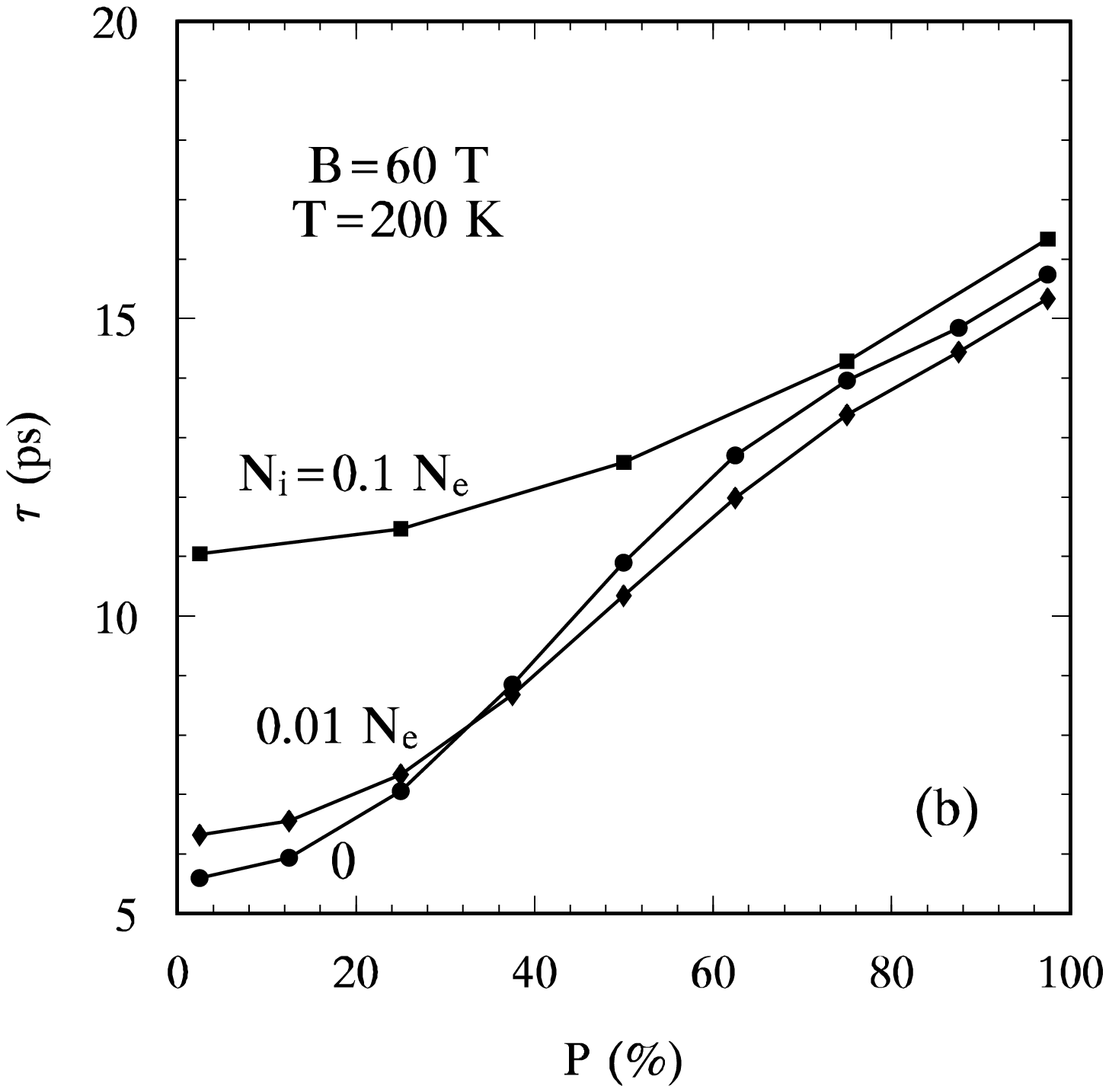,width=9.cm,height=8.5cm,angle=0}
  \caption{Spin dephasing time $\tau$ versus the initial spin
    polarization $P$ for a GaAs QW with different impurity levels. 
    Circle ($\bullet$): $N_i=0$;
    Diamond ($\blacklozenge$): $N_i=0.01N_e$;
    Square ($\blacksquare$): $N_i=0.1N_e$. The lines are plotted for
    the aid of eyes. 
  }
  \label{fig4} 
\end{figure}

The anomalous peak in the $\tau$-$P$ curve in low temperature region
originates from the electron-electron interaction, specifically the
Hartree-Fock (HF) self-energy [{\em i.e.}, the last terms in the
Eq. (\ref{eq:f_coh}) and (\ref{eq:rho_coh})].  If one removes the HF
term, the anomalous peak as well as the large increase of SDT
disappears.  It is pointed out in our previous paper that although the
HF term itself does not contribute to the spin dephasing
directly,\cite{wu_epjb_2000,wu_js_2001} it can alert the motion of the
electrons as 
it behaves as an effective magnetic field ${\bf B}^{\mbox{HF}}({\bf
k})$. Therefore, the HF term can affect the spin dephasing by
combining with the DP term.  For small spin polarization as commonly
discussed in the literature, the contribution of the HF term is
marginal. However, when the polarization gets higher, the HF
contribution becomes larger.  Especially the effective magnetic field
formed by the HF term contains a longitudinal component
[$B_z^{\mbox{HF}}({\bf k})$] which can effectively reduce the
``detuning'' of the spin-up and -down electrons, and thus strongly
reduces the spin dephasing, therefore the SDT increases with initial
spin polarization.\cite{c0210313}
Moreover, besides the initial polarization,
$\rho_{\bf k}$ and therefore $B^{\mbox{HF}}({\bf k})$
are also affected by the applied magnetic field. With higher magnetic
field, both gets larger.  Under the high magnetic field and when the
initial spin polarization reaches to a right value, the effective
magnetic field ${\bf B}^{\mbox{HF}}({\bf k})$ may reach the magnitude
comparable to the contribution from the DP term as well as the applied
magnetic field in the coherent parts of the Bloch equations and and
reduces the anisotropic caused by the DP term. Therefore, one gets
much longer SDT.  However, if one further increases the initial
polarization, the HF term exceeds the resonance condition. As the
result, the SDT decreases.  Therefore, one gets the anomalous peak
which is similar to the resonance effect. It is noted at, as both the
DP term and the HF term are $k$-dependent, the resonance is broadened.

For high temperatures the HF term is smaller.  In order to reach the
resonance, one needs to go to higher polarization. Therefore, as shown
in the figure the anomalous peak shifts to the higher
polarization. However, when the temperature is high enough, even
largest polarization $P=100$\ \% cannot make the HF term to reach
the resonance condition. Therefore, the peak disappears.

The $\tau$-$P$ curve is much different when the impurities are
introduced. It is seen from Fig.~\ref{fig3}(b) that, when the density
of impurity is large, say $N_i=0.1 N_e$, the fast rise in $\tau$-$P$
curve remains.  Nevertheless the increase is much smaller than the
corresponding one when the impurities are absent. In addition to the
reduction of the rise in $\tau$-$P$ curves, the impurities destroy the
anomaly too.  One can easily see that, with the impurity level
$N_i=0.1 N_e$, for all of the temperatures we study, the SDT increases
uniquely with the polarization.

To further reveal the contribution of the impurity to the dephasing
under different conditions, we plot the SDT as a function of the
polarization for different impurity levels at $T=120$\ K and 200\ K in
Fig.~\ref{fig4}(a) and (b) respectively.  The figure clearly shows
that for low temperature, the impurity tends to remove the anomalous
peak and to shift the peak to the larger initial spin
polarization. This is because that the impurity reduces the HF term
and therefore the resonance effect is also reduced. Hence, in order to
reach the maximum resonance, one has to increase the initial spin
polarization. Consequently, the peak shifts to larger $P$. Whereas
when $N_i$ is raised to $0.1 N_e$, the HF term is reduced too much to
form a peak.  It is also noted that for high temperatures, there is no
anomalous peak.

\begin{figure}[htb]
  \psfig{figure=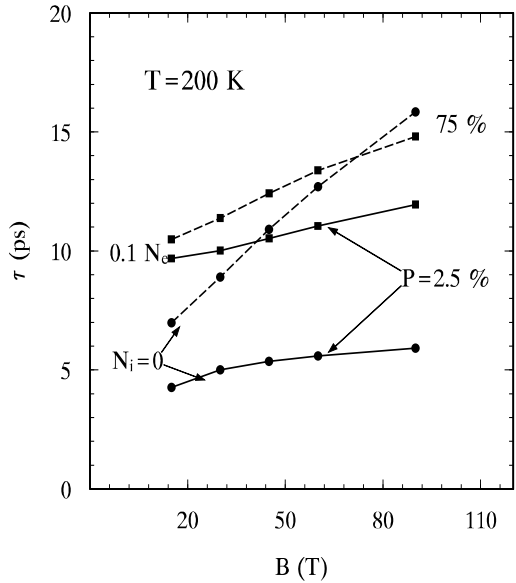,width=9.cm,height=8.5cm,angle=0}
  \caption{Spin dephasing time $\tau$ versus the applied magnetic
    field for GaAs QW's for different spin polarizations and different
    impurity levels. 
    Solid curve with circle dots: $N_i=0$, $P=2.5\ \%$; 
    Solid curve with square dots: $N_i=0.1N_e$, $P=2.5\ \%$; 
    Dashed curve with circle dots: $N_i=0$, $P=75\ \%$; 
    Dashed curve with square dots: $N_i=0.1N_e$, $P=75\ \%$.}
  \label{fig6}
\end{figure}

It is interesting to note that in the low polarized regime and when
the temperature is $120$~K, $\tau$ first decreases when the impurities
 are introduced. However, when we further increase the impurity level
 from $N_i=0.01N_e$ to $N_i=0.1N_e$, $\tau$ increases again.  As we
 pointed out before that the impurities affect the spin dephasing in
 two ways.\cite{wu_pss_2000} On one hand, the electron-impurity scattering
 provides a new spin dephasing channel through combining with the DP
 term\cite{DP,wu_pss_2000} to give an effective SF scattering and through the
 inhomogeneous broadening provided by the DP
 term.\cite{wu_pss_2000,wu_ssc_2002,wu_jpsj_2001,wu_epjb_2000,wu_js_2001} This gives rise to the
 enhancement of the dephasing.  On the other hand, the scattering also
 redistributes the electrons in the momentum space and leads them to an
 isotropic distribution. Therefore, the scattering can suppress the
 anisotropy caused by the DP term, consequently the effective
 spin-flipping scattering.  This leads to a smaller spin dephasing.
 Our result indicates that in low temperature and low polarization
 region, the impurities tend to reduce the SDT through the added spin
 dephasing channel when their concentration is small. When the impurity
 level increases, the impurities destroy the anisotropy introduced by
 DP effect more effectively and the electron-impurity scattering leads
 to an increase in the SDT.  This feature is different for high
 temperatures as shown in Fig.~\ref{fig3}(b), where the SDT increases
 with the increase of the impurity density. The resason is understood
 as the reduction of the inhomogeneous broadening for high temperature
 (see more detail in our previous paper Ref.\ \onlinecite{c0210313}). Therefore, the second effect
 mentioned above dominants. Consequently, the scattering tends to
 reduce the dephasing and the dephasing time increases with the
 concentration of the impurities.

\begin{figure}[htb]
  \psfig{figure=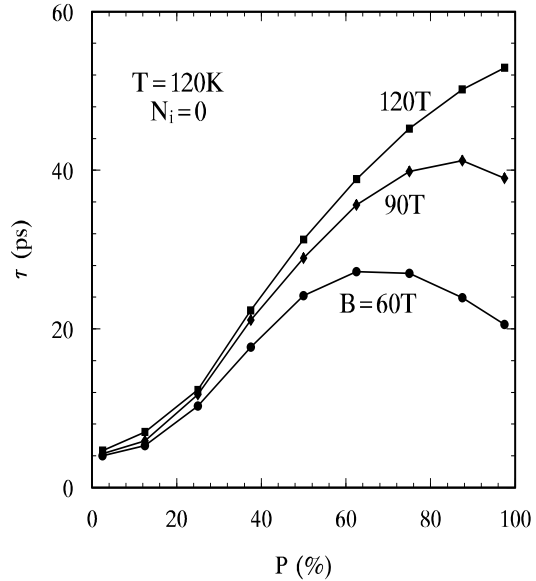,width=9.cm,height=8.5cm,angle=0}
  \caption{Spin dephasing time $\tau$ versus the polarization 
    for GaAs QW's at different magnetic field. 
    Circle ($\bullet$): B=60\ T; Diamond ($\blacklozenge$): B=90\ T; 
    Square ($\blacksquare$): B=120\ T. The lines are plotted for the aid
    of eyes.
  }
  \label{fig7}
\end{figure}

\subsection{Magnetic field dependence of the spin dephasing}

We now investigate the magnetic field dependence of the spin
dephasing.  In Fig.~\ref{fig6}, we plot the SDT versus the applied
magnetic field for different impurity levels and different spin
polarizations. It is seen that for all the cases we study, the SDT
increases with the magnetic field. This is because in the presence of
a magnetic field, the electron spins undergo a Larmor precession
around the magnetic field. This precession suppresses the precession
about the effective magnetic field ${\bf h}({\bf
k})$.\cite{meier,bronold} Therefor the SDT increases with the magnetic
field.  It is pointed out that in 3D electron gas, the magnetic field
also forces electrons to precess around it. This precession introduces
additional symmetry in the momentum space that limits the ${\bf
k}$-space available to the DP effect which is anisotropic in
it.\cite{meier,wu_pss_2000,bronold} This can further reduce the spin
dephasing.  However, it is expected that this effect in the 2D case is
less effective than the 3D case as in $z$-direction the momentum is
quantized and the momentum precession around the magnetic field should
be suppressed.

In additional to the above mentioned effect of the magnetic field on
spin dephasing, one can further see from Fig.~\ref{fig6}, that for
large polarization, the magnetic field also enhances the HF term. As
we mentioned before, for large polarization, the contribution from the
HF term is important. Increase of the HF term serves as additional
magnetic field which further suppresses the effect of the DP term
${\bf h}({\bf k})$, and therefore results in a faster rise in the
$\tau$-$B$ curve. To reveal more concrete about the combining effect
of the magnetic field and the HF term on spin dephasing, we plot the
SDT as a function of polarization in Fig.~\ref{fig7}. It is shown that
the rise in the $\tau$-$P$ curve increases with the magnetic field.
Moreover, the position of the peak in $\tau$-$P$ shifts to a larger
polarization. This is understood that, it needs a larger HF term, and
hence a larger spin polarization, to achieve the resonance condition
when the magnetic field increases.  When the magnetic field is raised
to 8\ T, it is no longer possible to form the resonance for all of the
polarization.  As a result the SDT increases uniquely with the
polarization and there is no peak in the $\tau$-$P$ curve.

\section{Conclusion}

In conclusion, we have performed a systematic investigation of the DP
effect on the spin dephasing of $n$-typed GaAs QW's under high 
magnetic fields in Voigt configuration. Based on the nonequilibrium
Green's function theory, we derived a set of kinetic Bloch equations
for a two-spin-band model. This model includes the electron-phonon,
electron-impurity scattering as well as the electron-electron
interaction. By numerically solving the kinetic Bloch equations, we
study the time evolution of electron densities in each spin band and
the spin coherence -- the correlation between spin-up and -down
bands. The SDT is calculated from the slope of the
envelope of the time evolution of the incoherently summed spin
coherence.  We therefore are able to study in detail how
this dephasing time is affected by various factors such as spin
polarization, temperature, impurity level, magnetic field and electron
density.  In this paper we focus ourselves on the special features of
the SDT under high magnetic fields. Features which are similar to
those in the moderate magnetic fields have been reported in our
previous paper Ref.\ \onlinecite{c0210313} and are therefore not repeated in this paper. 

It is discovered that the SDT increases with the initial spin
polarization.  Moreover, for low impurity level and low temperature,
there is a giant anomalous resonant peak in the curve of the SDT
versus the initial spin polarization.  This resonant peak moves to high
spin polarization and its magnitude is fast reduced (enhanced) until
the whole resonance disappears if one increases the impurity density
and/or the temperature (the magnetic field). It is discovered that
this anomalous resonance peak originates from the HF contribution of
the electron-electron Coulomb interaction.  Under the right spin
polarization, the contribution of HF term may reach the magnitude
comparable to the contribution of the DP term as well as the magnetic
field in the coherent parts of the Bloch equations
 and reduces the 
anisotropy caused by the DP effect---consequently reduces the spin
dephasing.  As the resonance is the combined effects of the HF term,
the DP term and the magnetic field, the magnitude and position of the
resonance peak are affected by all the factors that can alert the
magnitude of the HF term, such as temperature, impurity scattering,
magnetic field as well as the electron density: For a given impurity
concentration, when the temperature increases, the HF term
reduces. Consequently the $\tau$-$P$ curve is smoothed and the peak
position is moved to higher spin polarization; For impurities free
samples, if the the temperature is raised to 200$~K$, the HF term is
reduced too much to form a resonance and the anomalous peak
disappears; The same situation happens when the impurity level
increases at a given temperature as the scattering also lowers the HF
term. When the impurity level is raised to $0.1 N_e$ there is no
resonance in the temperature region we studied; While the increase of
the magnetic field enhances the HF term and results in a faster
increase of the SDT as well as a higher resonant peak in $\tau$-$P$
curve. Moreover, as the magnetic field becomes larger, it needs a
larger HF term and hence a larger polarization in order to achieve the
resonant condition. Therefore the peak position is also moved to
higher polarization. It is further noted that the resonance condition
can only be achieved with high magnetic fields. 
For moderate magnetic fields, the contribution of the magnetic field 
is not big enough to reach the resonance condition for
GaAs due to the small $g$ factor of this material.

\acknowledgments
MWW is supported by the  ``100 Person Project'' of Chinese Academy of
Sciences and Natural Science Foundation of China under Grant
No. 10247002. We would like to thank Mr. T. Rao for running the code for different parameters for us.

%% \bibliographystyle{apsrev}
%% \bibliography{ref}

\end {document}